\documentclass[10pt,conference, compsocconf, A4]{IEEEtran}

\usepackage{cite}
\usepackage[pdftex]{graphicx}
\usepackage{amsfonts,amssymb,amsmath,alltt}
\usepackage{array}
\usepackage{mdwmath}
\usepackage{mdwtab}
\usepackage{fixltx2e}
\usepackage{stfloats}
\usepackage{hyperref}
\usepackage{xspace}
\usepackage{color}
\usepackage{stmaryrd}

\newenvironment{ttbox}{\begin{alltt}\ttbraces\small\tt}%
                      {\end{alltt}}
\def\ttbraces{\let\.=\nobreak\chardef\{=`\{\chardef\}=`\}\chardef\|=`\\}

\newcommand\zb{\hspace*{-2em}\vspace*{-0.65cm}\begin{zed}}
\newcommand\ze{\end{zed}\vspace*{-0.65cm}}
\newcommand\bz{\vspace*{-0.65cm}\begin{zed}}
\newcommand\ez{\end{zed}\vspace*{-0.65cm}}

\newcommand{\red}[1]{{\textcolor{red}{#1}}}

\newcommand\tttimes{\mbox{\( \times \)}}

\begin{document}

\title{Designing Data Protection for GDPR Compliance into IoT Healthcare Systems}

\author{\IEEEauthorblockN{Florian Kamm\"uller}
\IEEEauthorblockA{Middlesex University London, UK\\
f.kammueller@mdx.ac.uk}
\and
\IEEEauthorblockN{Oladapo O. Ogunyanwo}
\IEEEauthorblockA{Middlesex University London, UK\\
OO896@live.mdx.ac.uk}
\and
\IEEEauthorblockN{Christian W. Probst}
\IEEEauthorblockA{Unitex Institute of Technology, NZ\\
cprobst@unitec.ac.nz}
}

\maketitle

\begin{abstract}
In this paper, we investigate the implications of the General Data Privacy Regulation (GDPR)
on the design of an IoT healthcare system. On 25th May 2018, the GDPR has become mandatory 
within the European Union and hence also for all suppliers of IT products. Infringements on 
the regulation are now fined with penalties of up 20 Million EUR or 4\% of the annual 
turnover of a company whichever is higher. This is a clear motivation for system designers 
to guarantee compliance to the GDPR. We propose a data labeling model  to support access 
control for privacy-critical patient data together with the Fusion/UML process to 
design GDPR compliant system. We illustrate this design process on the case study of IoT 
based monitoring of Alzheimer's patients that we work 
on in the CHIST-ERA project SUCCESS. 
\end{abstract}

\section{Introduction}
\label{sec:intro}
Infringements on 
the basic principles of data processing, rights of data subjects, or other non-compliances
to Articles of the GDPR are fined  with up to 20 million EUR or 4\% of the annual 
turnover of an undertaking whichever is higher (Article 79 (3a) \cite{gdpr:18}). 
Therefore, it is a crucial need for 
any company to find ways to achieve and keep GDPR compliance.

The General Data Privacy Regulation (GDPR) \cite{gdpr:18} is a 209 page legal
document. For small businesses, it might be hard to tackle such a complex
requirements specification.
For these reasons, we attempt in this paper,
to show practically how to overcome the difficulty of such a legal formulation by 
\begin{itemize}
\item[(a)] summarizing the legal text, 
highlighting the technically relevant parts and 
\item[(b)] providing a fairly generic application example taken from the IoT healthcare
context establishing a data protection model for private data and 
\item[(c)] using the Fusion/UML analysis and design process we produce a global 
          architecture supporting the requirements given by the GDPR. 
\end{itemize}
We use established techniques from security and software engineering to show how
the GDPR can be systematically mapped onto a formal system architecture specification.
Technically, we propose a combination of information flow control models for 
data protection in distributed systems 
by labeling data with security labels in the style of the Decentralized Label 
Model (DLM) \cite{ml:98} and employing the development Fusion/UML for analysis
and design of software systems. More precisely, we use an extended process \cite{bk:03}
that establishes a consistency relationship between analysis and design
and produces a formal specification for the implementation.

The contribution of this paper is to 
\begin{itemize}
\item break down the legal document of the
GDPR into more digestible technical requirements,
\item show how the Decentralized Labelling Model (DLM) can be 
integrated with the pragmatic software engineering 
methods Fusion/UML to produce a formally specified system architecture, 
\item illustrate the process on the SUCCESS IoT healthcare application.
\end{itemize}
We first give a brief overview of the General Data Protection Regulation (GDPR) 
pointing out which parts are relevant for technical design of information systems 
(Section \ref{sec:gdpr}). We then provide an overview of the background: after a 
short introduction to the extended Fusion/UML-translation process, we give an in 
depth summary and analysis of related work (Section \ref{sec:rel}).
In Section \ref{sec:app}, we introduce the application case study of our CHIST-ERA 
project SUCCESS \cite{suc:16} on security and privacy of IoT.   
We then show how parts of the GDPR data protection specification can be mapped to 
the Decentralized Data Label model (DLM). Next, we analyse the requirements of
the SUCCESS case study providing a system class model and operation schemas. A further 
design of object interactions finishes the system architecture.
The Fusion/UML-ObjectZ translation process \cite{bk:03} allows schematic translation
into a formal specification (Section \ref{sec:prop}). 
We conclude in Section \ref{sec:concl}.

\section{The European Standard GDPR}
\label{sec:gdpr}
The GDPR  (General Data Protection Regulation)
is in full called ``the regulation of the European parliament 
and the council on the protection of individuals with regard 
to the processing of personal data and on the free movement of 
such data''. For this paper, we use the final proposal \cite{gdpr:18}
as our source to provide a comprehensive summary of the main points 
relevant for a technical analysis. Despite the relatively large size of
the document of 209 pages, the relevant portion for this is only about 30 
pages (Pages 81--111, Chapters I to Chapter III, Section 3). 
Chapter I generally defines the scope of the regulation in terms of main purpose
(protection of individuals), material scope (personal data) and territories (in the Union).
%
Chapter II defines the principles for data processing and retention. 

While Chapters I and II provide essential definitions, more technical requirements for data
processing are provided in Chapter III, Sections 1 to 3.
\begin{itemize}
\item Section 1 describes Transparency and Modalities. Article 12 states that the
controller must provide any information and communication (specified in Article 14--20) 
``relating to the data subject in a concise transparent and intelligible and easily
accessible form \dots''.
\item Section 2 provides details of the access rights and the information that the controller 
must provide to a data subject on request (Articles 14, 14a, 15), like the retention time and 
the purpose of data collection. 
\item Section 3 defines the right of a data subject to rectification and erasure of personal
data (``right to be forgotten'') as well as the right to restrict its processing (Articles 15--18).
\end{itemize}
In summary, Chapter III specifies that the controller must give the data subject 
{\it read access} (1) to any information, communications, and ``meta-data'' of the 
data, e.g., retention time and purpose. In addition, the system must enable 
{\it deletion of data} (2) and restriction of processing. 

An invariant condition for data processing resulting from these Articles 
is that the system {\it functions} must {\it preserve} any of the access 
rights of personal data (3).

\section{Background}
\label{sec:rel}

\subsection{The Extended Fusion/UML Software Development Method}
The method Fusion/UML is a software development process comprising
analysis and design phases. It defines consistency rules enabling the
control of consistency {\it in} the various models and {\it in between} them
throughout the development process. 
The final result of the Fusion/UML process is an accurate description of the
class interfaces suitable for any object oriented programming language.
The analysis phase is concerned with the question: {\it what} is and does
the system. 
One major result of this first step of Fusion/UML is a so-called {\it
operation model} describing in schematic form the system operations. 
The second part, the design, is concerned with the question: {\it how}
does the system achieve its goals. One major model of the design is the
so-called {\it object interaction model}. It uses UML collaborations to
define the system operations by message flows of method invocations between
objects in the system. Analysis and design contain various other models, but
the two parts operation model and object interaction model contain most of
the information about the operational part of the system design.
The paper \cite{bk:03} extends the Fusion/UML by a 
systematic translation of the operation model and the object interaction
model into the formal specification language Object-Z \cite{DR00,S00}. 
Object-Z is suitable as it constitutes an extension of Z by
the concepts of classes and objects and has a well established reference
semantics \cite{SKS02}.

In the context of Security and Privacy for the IoT for the SUCCESS project, 
this method is applied to systematically derive a formal specification
for the software system managing the secure data handling between the 
distributed entities. The challenge lies in the security part which poses
special requirements as put up by the GDPR but also more generally due
to the difficult nature of security as a non-functional requirement.
However, for the same reasons security needs to be designed into the system
from the beginning. If cannot be ``plugged onto'' a system at later stages --
an additional motivation to use a well-established software engineering method 
like Fusion/UML.

\subsection{Related Work on Privacy in IoT Healthcare systems}
A mobile application solution for healthcare called 
the Electro Cardiogram Android App (ECG App) \cite{jtola:14}
allows the end user to view data logging functionalities and ECG waves in 
the background. The application 
allows logged data to be uploaded to either a specific medical cloud, or the user's 
private centralized cloud, here the healthcare and patient monitored records are kept 
and can also be retrieved and viewed for analysis by medical personnel.
The Design of the entire system of the proposed solution in this paper is based on a 
layered architectural design pattern that divides the system into 3 different units 
called layers; the hardware layer which contains the IOIO microcontroller and sensors 
to collect signal data; the application layer which receives signal data sent from the 
hardware layer, and also contains 3 sub layers; and the Cloud layer which has the FTP 
server present in it, and receives the file from the FTP Client. The Server is also 
responsible for storing file in the File Table Technology.
The ECG Mobile app makes use of an IOIO microcontroller board, that obtains signal 
from a person using ECG electrodes and sends to the mobile device wirelessly using 
Bluetooth technology. The monitored ECG waves of a patient displayed in the mobile app 
is stored in Binary format which will be encrypted and uploaded to an SQL server private 
database in a secure manner making use of the FTPES protocol.
This System though tested however is only limited to monitoring ECG waves, but with 
proposed design improvements to be applied to other medical applications. 

In order to better protect and preserve sensitive patient health information, 
the paper \cite{erb:16} 
 proposes a holistic privacy middleware which they called ECMP middleware, that executes 
a two-stage concealment process within a distributed data protection protocol that 
utilizes the hierarchical nature of IOHT devices. This proposed solution complies with 
the Organization of Economic Cooperation and Development (OECD) privacy principles.
The OECD privacy principles are a set of fair information practice that can be considered 
as the primary components that enables the protection and privacy of personal data for 
cloud based services. The principles covers; Collection limitation, Data quality, 
Purpose specification, Use limitation, Security safeguard, Openness, Individual 
participation, and Accountability.
The ECMP is the main architectural element of the framework because it is responsible 
for the execution of the topological formation protocol for data collection. The ECMP 
middleware is deployed in 3 usage layers; the data collection layer which consists of 
the various IOHT devices and the mobile application that is used to capture the vital 
signs from the user's body or home , and also collects related data external. This data 
is used to make recommendations and treatment; the intermediate layer that consists of 
2 fog-nodes, where each of the nodes host the holistic privacy middleware that will execute 
the second stage of the two-stage concealment; the final layer is the service layer that 
consists of the various healthcare web services that are hosted on the cloud platforms. 
It also facilitates the collaboration between various service providers. 

The paper \cite{btv:17} 
introduces a smart gateway that safeguards the entire healthcare system using a modified 
Host Identity Protocol Diet Exchange (HIP-DEX) key exchange protocol and a new key exchange 
scheme based on Low Energy Adaptive Clustering Hierarchy (LEACH) routing protocol. For the 
purpose of this research demonstration, the health monitoring of sports personnel was 
selected. The research security approach is based on a lightweight mutual verification 
and key exchange protocol which is based on a hash function with no restriction on data 
storage, operation speed, size of data input etc. Using a lightweight modified HIP-DEX 
key exchange scheme, a secure link can be created between the cloud and the end-user device.
The entire architecture of the proposed system is based on a secure gateway which is 
implemented using Arduino MKRzero microcontroller board. After the parameters of the 
sports person has been collected using heartbeat/pulse, muscle and blood pressure sensors, 
the sensors collect the data and communicate using a pre-shared key that can be specified 
during the device configuration time. The gateway then analyses the received data for any 
abnormalities in it using an adaptive rule engine. In the case that abnormalities are 
discovered, the monitoring device will be notified using a fast and secure channel enabled
using a new key exchange scheme based on the LEACH protocol. The channel between the 
gateway and the cloud is modified using the HIP-DEX protocol. End users such as the 
doctors, nurses and other authorized personnel can frequently view and monitor the 
health statistics using an android mobile application.

\section{Application Example from IoT Healthcare}
\label{sec:app}

\subsection{System Model}
\begin{figure}[h]
\includegraphics[scale=.19]{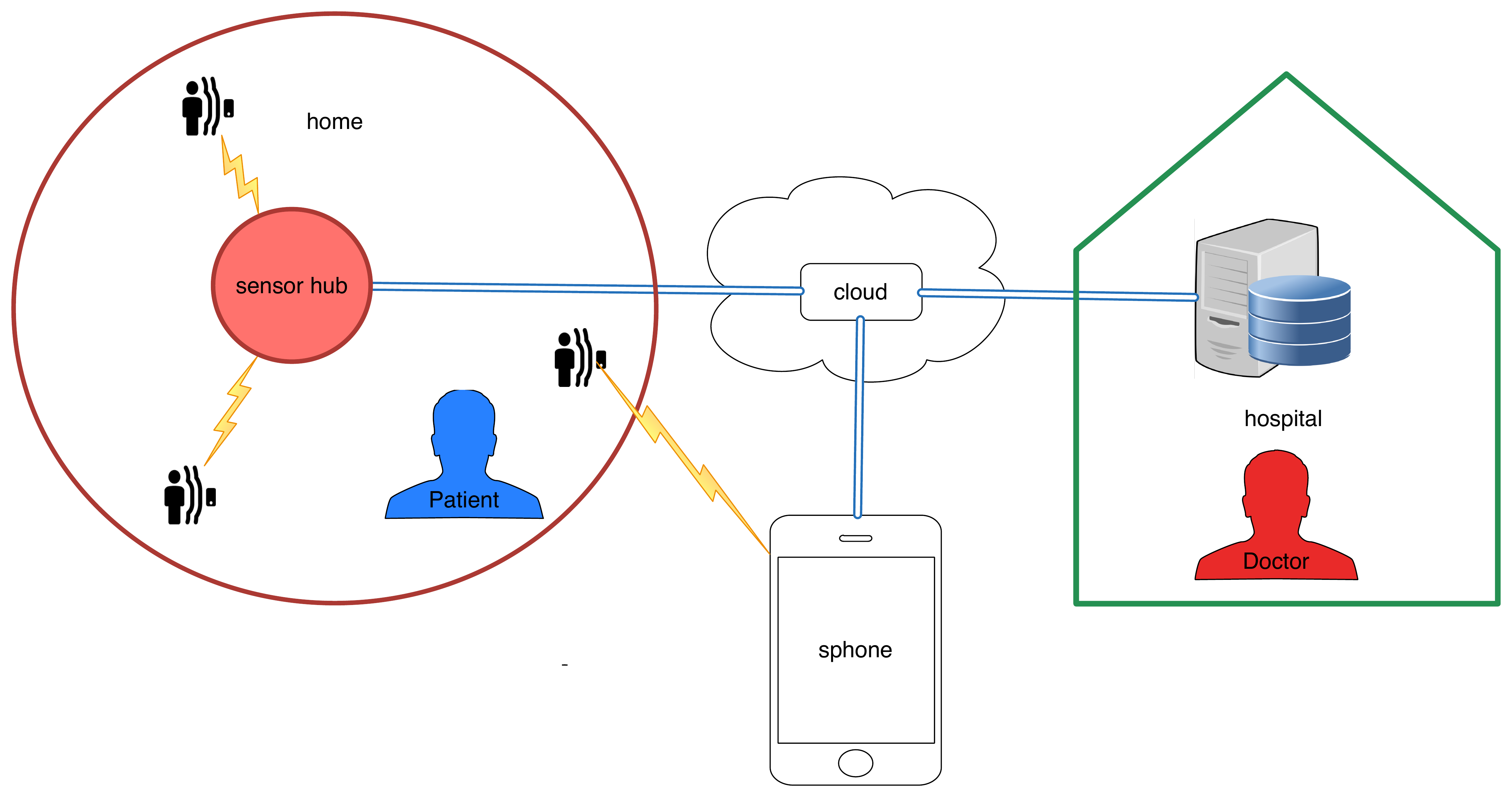}
\caption{IoT healthcare monitoring system for SUCCESS project}\label{fig:iot}
\end{figure}

The example of an IoT healthcare systems is from the CHIST-ERA project SUCCESS \cite{suc:16}
on monitoring Alzheimer's patients. 
Figure \ref{fig:iot} illustrates the system architecture where data collected by sensors 
in the home or via a smart phone helps monitoring bio markers of the patient. The data 
collection is in a cloud based server to enable hospitals (or scientific institutions) 
to access the data which is controlled via the smart phone.

Before embarking on an in depth security analysis and design, we observe
a few issues related to the context and main stream technologies which 
we have to put up with:
\begin{itemize}
\item Sensors and sensor hub hardware run proprietary operating systems and
   are thus outside of the control we have over such systems. A plethora of
   attacks can easily be found and more be imagined. 
\item The Privacy sensitivity of the sensored data is difficult to measure.
   For example, the fact that the patient walks into a room, as observed by 
   a motion sensor, is not critical. However, if the behavioural algorithm that
   accumulates motion data into the data ``patient is wandering'' -- which 
   is an indicator for early Alzheimer's symptoms, the data becomes privacy critical.
\end{itemize}
Therefore, we exclude the home 
from the security perimeter. We stipulate this as a necessary security assumption.
This is the best we can do. Consequently, data cannot be held in the server in 
the home but needs to be uploaded immediately to the cloud server. Similarly, 
sensor data transmitted via the smartphone is not kept on the phone. 

Within the security perimeter, we thus place only the cloud server and the connected
hospital (or other client institutions). The smartphone and the home server feature
as data upload devices and the smartphone additionally as a control device that is
included in some of the use cases.

\begin{figure*}[ht!]
\begin{center}
\setlength{\unitlength}{0.15mm}
{\small
\begin{picture}(750,150)
\thicklines
\put(0,120){\framebox(120,60){\footnotesize $\begin{array}{l}{{\sf home}}\\
                                                 \    \end{array}$}
}
\put(0,150){\line(1,0){120}}
\put(120,140){\line(1,0){160}}
\put(130,145){$*$}
\put(260,145){\footnotesize$1$}

\thicklines
\put(0,0){\framebox(120,60){\footnotesize $\begin{array}{l}{\small {\sf sphone}}\\[1ex]
                                                 {\sf PIN} \end{array}$}
}
\put(0,35){\line(1,0){120}}
\put(120,45){\line(1,0){120}}
\put(240,45){\line(0,1){75}}
\put(240,120){\line(1,0){40}}
\put(130,50){$*$}
\put(260,100){\footnotesize$1$}

\thinlines
\put(200,-30){\framebox(280,240)}
\put(210,220){{\rm ${\sf system\ border}$}}
\red{\put(195,-35){\framebox(500,250)}}
\put(500,220){{\rm ${\sf \red{security\ perimeter}}$}}

\thicklines
\put(280,100){\framebox(120,100){\footnotesize $\begin{array}{l}{\sf Auth}\\[.6ex]
                                                 {\sf patients} \\[.6ex]
                                                 {\sf reg\_usrs}\end{array}$}}

\put(280,165){\line(1,0){120}}
\put(400,140){\line(1,0){130}}
\put(410,145){\footnotesize$1$}
\put(450,145){\footnotesize ${\sf Has}$}
\put(510,145){$*$}
\thicklines
\put(280,-20){\framebox(120,60){\footnotesize $\begin{array}{l}{\sf DB}\\[.6ex]
                                                {\sf table} \end{array}$}}
\put(280,15){\line(1,0){120}}
\put(340,40){\line(0,1){60}}
\put(325,45){\footnotesize$1$}
\put(325,85){\footnotesize$1$}
\put(345,65){\footnotesize${\sf Controls}$}
\put(530,100){\framebox(120,100){\footnotesize $\begin{array}{l}{\sf hospital}\\[.6ex]
                                                 {\sf staff}\\[.6ex]
                                                 {\sf table}\end{array}$}}
\put(530,165){\line(1,0){120}}

\end{picture}
}
\end{center}
\caption[]{System class model for IoT healthcare system 
}\label{fig:arch}
\end{figure*}
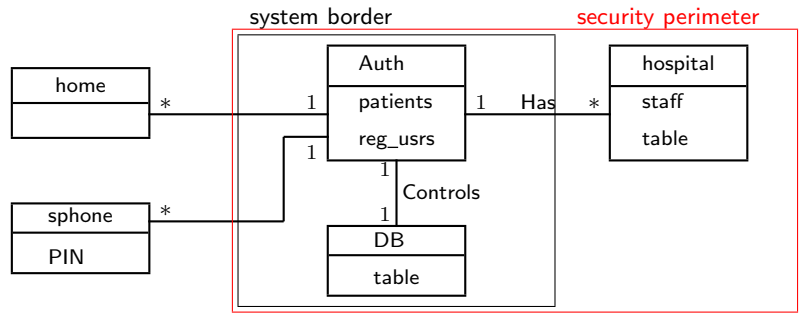

We cut short the full Fusion/UML process and present just its outcome corresponding
to the analysis process based on these observations. This first outcome, 
the Fusion/UML analysis model, defines the system class model as depicted in 
Figure \ref{fig:arch}.

Another result of the Fusion/UML analysis along with this system architecture 
is a set of operations schemas based on the system class model and additional
use cases. We present them together with the system design using object collaborations
in the following section.

\section{A Global GDPR compliant IoT Architecture}
\label{sec:prop}

The Decentralised Label Model (DLM) \cite{ml:98} introduced the idea to
label data by owners and readers. We pick up this idea but extend labels with
additional information, like purpose and retention time, to cover all requirements 
of the  GDPR data protection requirements. 

In this section, we first introduce the necessary details about DLM and concepts
for extending this model for our purposes. Then we present the use cases for our
IoT healthcare application as a suite of system operations defined by operation schemata 
and designed into object interactions. Using the system class model from the previous 
section, we can apply the extended Fusion/UML process \cite{bk:03} and derive a formal
system specification.

\subsection{Security and Privacy by Labeling Data}
\label{sec:label}
We firstly need to specify the owner and the set of readers given by the
following type \texttt{dlm}.
\begin{ttbox}
type dlm = actor \tttimes actor set
\end{ttbox}
Labelled data is then just given by the type \texttt{dlm \tttimes\ data}
where \texttt{data} can be any data type. Additional meta-data, like retention
time and purpose, can be encoded as part of this type \texttt{data}. We omit
these detail here for conciseness of the exposition.
We may use the concept of erasure \cite{hs:08} to implement the latter.

Using labeled data, we can now express the essence of Article 4
Paragraph (1): 'personal data' means any information relating to an 
identified or identifiable natural person ('data subject').

\subsection{Use Cases}
The Fusion/UML analysis actually produces the system class model with
the system border only after the use case model has been specified. 
However, we take the liberty to present them only now since we needed 
to introduce DLM first and can now better motivate the decision of 
putting the extended system border as a security perimeter to include
the hospital server.

There are four use cases for the IoT healthcare application:
\begin{enumerate}
\item User data is uploaded to cloud from home server or via mobile phone.
\item User wants to delete or change data using his mobile phone as control device.
\item Others (hospital, researchers, GPs) access data in the cloud.
\item System needs to maintain distributed consistency and therefore
\begin{enumerate}
     \item ensure that only label preserving operations are allowed;
     \item data is tracked within the system;
     \item time (and erasure) is monitored and enforced on all data in the system.
\end{enumerate}
\end{enumerate}

In the previous section, we already defined the system border as a security perimeter 
that encompasses not only the access controlled cloud data base but also the hospital 
server into one system. Consequently, the security labels must be maintained 
consistently by the hospital and thus within the healthcare system\footnote{Remember, 
we excluded the home server from the system as a necessary security assumption in 
the previous section.}.
This design decision is now justified by Use Case 4.

\subsection{Operation Schemata}
Although in their particular incarnation novel to the 
UML, operation schemata are just a composition of UML tagged values. 
The values tagged together are: {\bf Operation, Description, Inputs, Reads,
Changes, Sends, Pre}-condition, and {\bf Post}-condition.
That is, an operation schema contains the operation's name, its informal
description, and inputs. Furthermore, such a schema defines the objects and
associations from which the operation reads and on which it writes, further
specifying concrete conditions on those objects using a {\bf with}-clause
followed by a formal condition. The pre- and postconditions tagged with
an operation schema may define formal conditions about the operation.

The operation schema for the upload operation defines the system
function that is required by Use Case 1: a data item d and its
intended DLM-label (o,r) are input by an actor who further provides
an id as its identity. Provided that this id is the same as the inputting 
actor's identity o, the database is updated and a message about the 
successful completion is sent to the initiator. The uploading party could
be the smartphone or the home server expressed by the set of classes \{home,sphone\}.

\vspace{.1cm}

\noindent%
{\small
\begin{tabular}{|lcp{5.8cm}|}\hline
{\bf Operation} & = &  upload\\
{\bf Description} & = &    The home server or mobile phone uploads
                    patient data to the cloud server.
\\\hline
{\bf Input} & = & d: data, (o,r): dlm, id: actor \\
{\bf Reads} & = &  {\underline {as: Auth}} {\bf with} o $\in$ as.patients $\wedge$ r $\subseteq$ as.reg\_usrs, Controls\\ 
{\bf Changes} & = & {\underline{db: DB}} {\bf with} (as,db) $\in$ Controls\\
{\bf Sends} & = & {\underline{:\{home,sphone\}}}:\{upload\_ok\} \\
{\bf Pre} & = & id = o \\
{\bf Post} & = & db.table' = db.table $\cup$ \{((o,r),d)\} $\wedge$\\
           & & {\bf is\_sent}\{trans\_ok\}
\\\hline 
\end{tabular}}

\vspace{.1cm}

The operation schema for delete is very similar in its structure to upload but
it can only be initiated from the smart phone which servers as the control unit
for the user. If the data item with the right dlm label as input by the user
exists, it is deleted from the database {\bf and} hospital servers.

\vspace{.1cm}

\noindent%
{\small
\begin{tabular}{|lcp{5.8cm}|}\hline
{\bf Operation} & = &  delete\\
{\bf Description} & = &    The mobile phone deletes
                    patient data on the cloud server and hospital.
\\\hline
{\bf Input} & = & d: data, (o,r): dlm, id: actor \\
{\bf Reads} & = &  {\underline {as: Auth}} {\bf with} o $\in$ as.patients $\wedge$ r $\subseteq$ as.reg\_usrs, Controls\\ 
{\bf Changes} & = & {\underline{db: DB}} {\bf with} (as,db) $\in$ Controls\\
              &   & {\underline{h: Hospital}} {\bf with} (as,h) $\in$ Has\\
{\bf Sends} & = & {\underline{:\{sphone\}}}:\{delete\_ok\} \\
{\bf Pre} & = & id = o $\wedge$ ((o,r),d) $\in$ db.table $\wedge$\\
          &   &  ((o,r),d) $\in$ h.table\\
{\bf Post} & = & db.table' = db.table $\setminus$ \{((o,r),d)\} $\wedge$\\
           &  & h.table' = h.table $\setminus$ \{((o,r),d)\} $\wedge$\\
           &  & {\bf is\_sent}\{delete\_ok\}
\\\hline 
\end{tabular}}

\vspace{.1cm}

The operation schema for download is, contrary to the previous two operations,
initiated from the hospital side: on input of an id (of a staff member of the hospital),
a data item d is copied over to the hospital's data table if the DLM label permits
this, that is, the hospital h is named in the reader labels r. Besides the message
``download\_ok'' a second message ``access'' is sent to the user's smart phone to inform
of the data access.

\vspace{.1cm}

\noindent%
{\small
\begin{tabular}{|lcp{5.8cm}|}\hline
{\bf Operation} & = &  download\\
{\bf Description} & = &    A doctor downloads patient data from the 
  cloud server to the hospital server.
\\\hline
{\bf Input} & = & o, id: actor \\
{\bf Reads} & = &  {\underline {as: Auth}} {\bf with} o $\in$ as.patients $\wedge$ r $\subseteq$ as.reg\_usrs,
                    {\underline {db: DB}} {\bf with} (as,db) $\in$ Controls\\ 
{\bf Changes} & = & {\underline{h: Hospital}} {\bf with} (as,h) $\in$ Has\\
{\bf Sends} & = & {\underline{:\{sphone\}}}:\{access\}, \\
            &   & {\underline{:\{doctor\}}}: \{download\_ok\} \\
{\bf Pre} & = & ((o,r),d) $\in$ db.table $\wedge$ id $\in$ h.staff $\wedge$ h $\in$ r $\wedge $\\
          &   &  o $\in$ as.patients $\wedge$ h $\in$ as.reg\_usrs\\
{\bf Post} & = & h.table' = h.table $\cup$ \{((o,r),d)\} $\wedge$\\
           &  & {\bf is\_sent}\{download\_ok\} $\wedge$ {\bf is\_sent}\{access\} 
\\\hline 
\end{tabular}}

\subsection{Design: Object Collaborations}
The system design is the phase of the software development process that follows
the analysis, and needs to be consistent with it. This implies that for all
system operations defined in the analysis, the {\it object interaction
model} has to define {\it object interaction graphs} describing the
execution of the system operation on the objects in the system. 
To that end, the object interaction graphs introduce new operations: the
graphs are UML collaborations. They define message flows in a sequential
order using numbers. The object interaction starts from an actor executing the system
operation as a message to an object called the {\it controller} of that system
operation. The controller delegates the initial system operation to
so-called {\it collaborators}, objects of the system that are associated to
the controller. The initial task set out by the controller can be 
delegated in turn by the collaborators to other associated objects.
All objects that are corresponding with each other must be connected by associations in
the system class model.  This is a consistency conditions. Others are given
by the selection, pre- and postconditions defined in the system operations.
UML annotation with $\{ \ \}$ may be employed to
define the selection of this particular object or annotating pre- or
postcondition for the method calls contained in the messages tagging
it to the objects.

\begin{figure*}[ht!]
\begin{center}
\setlength{\unitlength}{0.15mm}
{\small
\begin{picture}(900,150)
\thicklines
\put(0,80){\line(1,1){25}}
\put(25,105){\line(1,-1){25}}
\put(25,105){\line(0,1){50}}
\put(0,145){\line(1,0){50}}
\put(25,165){\circle{20}}
\put(-5,60){\small ${\sf \underline{x: \{home, sphone\}}}$}
\put(80,130){\line(1,0){200}}
\put(90,140){\small ${\sf upload(id,((o,r),d))}$}
\thinlines
\put(120,165){\vector(1,0){80}}

\thicklines
\put(280,100){\framebox(120,60){\small$\sf \underline{as: Auth}$}}
\put(350,75){$\{ {\sf x = id = o}\ \wedge $}
\put(350,50){$\ {\sf o} \in {\sf as.patients}\ \wedge$}
\put(350,25){$\ {\sf r} \subseteq {\sf as.reg\_usrs} \}$}

\put(400,130){\line(1,0){200}}
\put(420,140){\small$ \sf 1: add(((o,r),d))$}
\thinlines
\put(450,160){\vector(1,0){95}}

\thicklines
\put(600,100){\framebox(120,60){\small$\sf \underline{db:DB}$}}

\put(650,75){$\{ {\sf db.table' = db.table} \cup \{{\sf ((o,r),d)}\} \}$}

\put(340,100){\line(0,-1){100}}
\put(340,0){\line(-1,0){315}}
\put(25,0){\line(0,1){50}}
\put(120,10){2: \small ${\bf send\ to}\{{\sf upload\_ok}\}$}
\thinlines
\put(300,30){\vector(-1,0){140}}
\end{picture}
}
\end{center}
\caption{Object collaboration diagram for the upload.}\label{fig:OIupload}
\end{figure*}
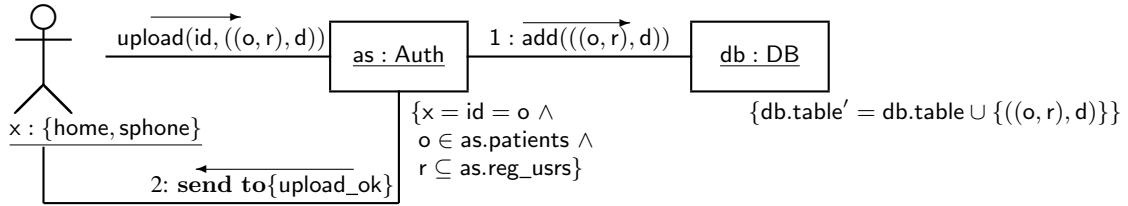

The operation for upload leads to the object collaboration illustrated in 
Figure \ref{fig:OIupload}. The preconditions already identified 
in the system operation must hold, that is, the initiator's id must match the
owner label o, must be a registered patient, and the reader label set r must only
contain registered readers.  
Since we now look at actual objects and not classes
(like in the analysis), we can additionally identify an actual object x of either
class home or sphone as the initiating object and further require that this
object corresponds to the actor o and id, respectively. This specifies authentication
and must be implemented by some authentication protocol in an implementation.
In the internal step 1, the system operation upload delegates the data 
upload to the collaborator object db of class DB and tags the postcondition
representing the change of the data base as a UML annotation tagged to db.

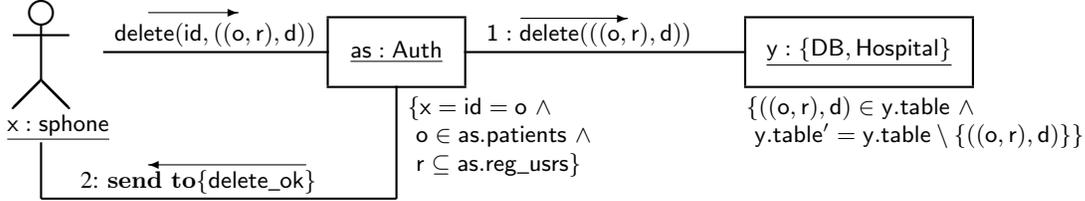
\begin{figure*}
\begin{center}
\setlength{\unitlength}{0.15mm}
{\small
\begin{picture}(1000,150)
\thicklines
\put(0,80){\line(1,1){25}}
\put(25,105){\line(1,-1){25}}
\put(25,105){\line(0,1){50}}
\put(0,145){\line(1,0){50}}
\put(25,165){\circle{20}}
\put(-5,60){\small ${\sf \underline{x: sphone}}$}
\put(80,130){\line(1,0){200}}
\put(90,140){\small ${\sf delete(id,((o,r),d))}$}
\thinlines
\put(120,165){\vector(1,0){80}}

\thicklines
\put(280,100){\framebox(120,60){\small$\sf \underline{as: Auth}$}}
\put(350,75){$\{ {\sf x = id = o}\ \wedge $}
\put(350,50){$\ {\sf o} \in {\sf as.patients}\ \wedge$}
\put(350,25){$\ {\sf r} \subseteq {\sf as.reg\_usrs} \}$}

\put(400,130){\line(1,0){250}}
\put(420,140){\small$ \sf 1: delete(((o,r),d))$}
\thinlines
\put(450,160){\vector(1,0){95}}

\thicklines
\put(650,100){\framebox(200,60){\small$\sf \underline{y:\{DB, Hospital\}}$}}

\put(650,75){$\{ {\sf ((o,r),d) \in y.table}\ \wedge $}
\put(650,50){$ \ {\sf  y.table' = y.table \setminus \{{\sf ((o,r),d)}}\} \}$}

\put(340,100){\line(0,-1){100}}
\put(340,0){\line(-1,0){315}}
\put(25,0){\line(0,1){50}}
\put(60,10){2: \small ${\bf send\ to}\{{\sf delete\_ok}\}$}
\thinlines
\put(260,30){\vector(-1,0){140}}
\end{picture}
}
\end{center}
\caption{Object collaboration for delete operation.}\label{fig:OIdelete}
\end{figure*}

The operation for delete leads to the object collaboration illustrated in Figure \ref{fig:OIdelete}.
Provided the conditions similar to the previous case and consistent with the operation schema hold,
the system operation delete is simultaneously delegated to corresponding methods in data base 
objects and hospital objects. 
In the collaboration diagram, the class set \{DB, Hospital\} generalises over the two 
branches one for the data base and one for hospitals. This simplifies the diagram but is
just a graphical abbreviation.

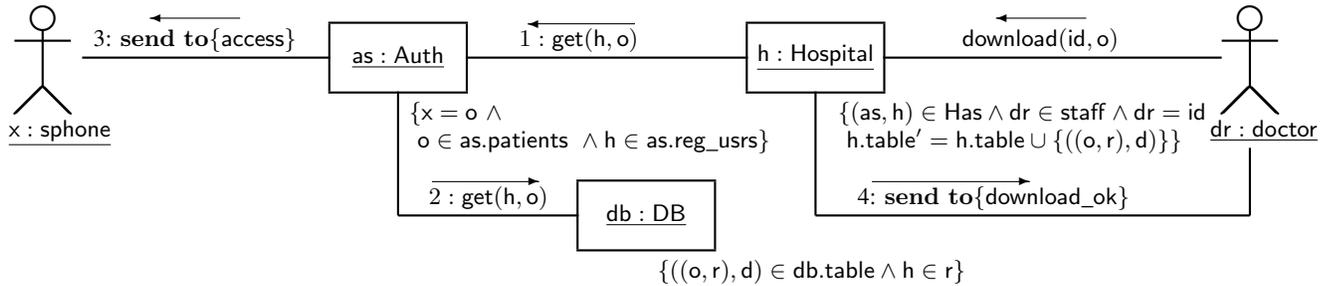
\begin{figure*}
\begin{center}
\setlength{\unitlength}{0.15mm}
{\small
\begin{picture}(1100,190)
\thicklines
\put(0,120){\line(1,1){25}}
\put(25,145){\line(1,-1){25}}
\put(25,145){\line(0,1){50}}
\put(0,185){\line(1,0){50}}
\put(25,205){\circle{20}}
\put(-5,100){\small ${\sf \underline{x: sphone}}$}
\put(60,170){\line(1,0){220}}
\put(70,180){\small 3: ${\bf send\ to}\{{\sf access}\}$}
\thinlines
\put(200,205){\vector(-1,0){80}}

\thicklines
\put(280,140){\framebox(120,60){\small$\sf \underline{as: Auth}$}}
\put(350,115){$\{ {\sf x = o}\ \wedge $}
\put(350,90){$\ {\sf o} \in {\sf as.patients}\ \wedge {\sf h} \in {\sf as.reg\_usrs} \}$}

\put(400,170){\line(1,0){250}}
\put(440,180){\small\ $ 1: {\sf get(h,o)}$}
\thinlines
\put(550,200){\vector(-1,0){95}}

\thicklines

\put(500,0){\framebox(120,60){\small$\sf \underline{db:DB}$}}
\put(570,-25){$\{ {\sf ((o,r),d) \in db.table \wedge h \in r }\}$}
\put(340,140){\line(0,-1){110}}
\put(340,30){\line(1,0){160}}
\put(360,40){\small\ $ 2: {\sf get(h,o)}$}
\thinlines
\put(370,60){\vector(1,0){95}}

\thicklines
\put(650,140){\framebox(120,60){\small$\sf \underline{h:Hospital}$}}

\put(730,115){$\{  {\sf (as,h) \in Has \wedge dr \in staff \wedge dr = id}$}
\put(730,90){$\ {\sf h.table' = h.table \cup \{((o,r),d) \}} \} $}

\put(710,140){\line(0,-1){110}}
\put(710,30){\line(1,0){385}}
\put(1095,30){\line(0,1){60}}
\put(740,40){ 4: \small ${\bf send\ to}\{{\sf download\_ok}\}$}
\thinlines
\put(760,60){\vector(1,0){140}}

\thicklines
\put(1070,120){\line(1,1){25}}
\put(1095,145){\line(1,-1){25}}
\put(1095,145){\line(0,1){50}}
\put(1070,185){\line(1,0){50}}
\put(1095,205){\circle{20}}
\put(1060,100){\small ${\sf \underline{dr: doctor}}$}
\put(770,170){\line(1,0){300}}
\put(840,180){\small ${\sf download(id,o)}$}
\thinlines
\put(950,205){\vector(-1,0){80}}

\end{picture}
}
\end{center}
\caption{Object collaboration for the download operation.}\label{fig:OIdownload}
\end{figure*}

The operation for download leads to the object collaboration illustrated in Figure \ref{fig:OIdownload}.
The collaboration graph nicely shows the two sides hospital and smart phone involved by having two 
actors included. Conditions are again directed mainly to ascertain authenticity and 
label correctness, and then a two-level delegation leads to the cloud database.
For conciseness, we only define the two get-method calls. Clearly -- and this is specified
in the UML-tag condition h.table' = h.table $\cup$ \{((o,r), d)\} -- the calls to get
lead to the retrieved labelled data d to be copied inversely up to hospital. However,
following the spirit of Fusion/UML, we omit that level of implementation detail in 
the specification.

The object collaborations nearly finalize the Fusion/UML process. The developed models
of analysis and design can then be schematically transformed into a set of interface 
specifications. In the extended Fusion/UML process, a schematic translation process
produces a complete Object-Z specification.
We omit this output here for reasons of space but it is attached as an appendix.

\section{Conclusions}
\label{sec:concl}
In this paper, we have summarised the new General Data Protection Regulation
(GDPR) and illustrated on an IoT healthcare patient monitoring system, how to 
design a system architecture specifying the privacy access control using the 
decentralized label model (DLM) and employing the extended Fusion/UML method
as software development process.

We have in detail discussed related work for IoT healthcare systems in Section
\ref{sec:rel}. Although some of these works address data protection, none does 
specifically address GDPR. 

The results of applying the Fusion/UML method to Security and Privacy of IoT are
generally interesting. Firstly, a security argument at the early requirements stage
shows that security protection in the home is for the application of a home based
IoT patient monitoring system futile. Therefore, we restricted our attention to
the cloud and hospital server part of the system. For the development of a system
design specification, the Fusion/UML method appeared to be working well.

There are two main observations of some importance. 
In the application, we have realized that we needed to consider the system distributed
over the cloud and hospital server as one system to enforce consistent labeling.
This security we simply emulated by abusing the system class border as a ``security perimeter''
pretending it to be one system. This could be more explicitly addressed as a 
feature in Fusion/UML and also in implementations (for example, using a distributed
ledger, a blockchain, to enforce consistent labeling across distributed servers).
Secondly, there is the recurring issue of authentication. In the application, we have
augmented that by explicitly stating identity of real and communicated actor identities,
for example, id = o. This could also be more explicitly addressed by using a dedicated
notation like auth(id,o).
It would be a valuable continuation of this initial experiment, to extract
these two major observations and refine the (extended) Fusion/UML method to 
accommodate them to provide a Security enhanced Fusion/UML.

Object-Z as a target language could be replaced by other
formal modeling frameworks potentially better suited for security analysis.
In a dedicated tool for security protocols, like ProVerif, 
it might be impossible to embed all of the abstract notions of object-oriented 
systems. Another possibility would be to use a more expressive framework, for example,
the Isabelle Insider framework \cite{kp:16}.


\appendix
\section{Translation into Object-Z}
\label{sec:app}
\begin{figure*}
\begin{center}
\includegraphics[scale=.7]{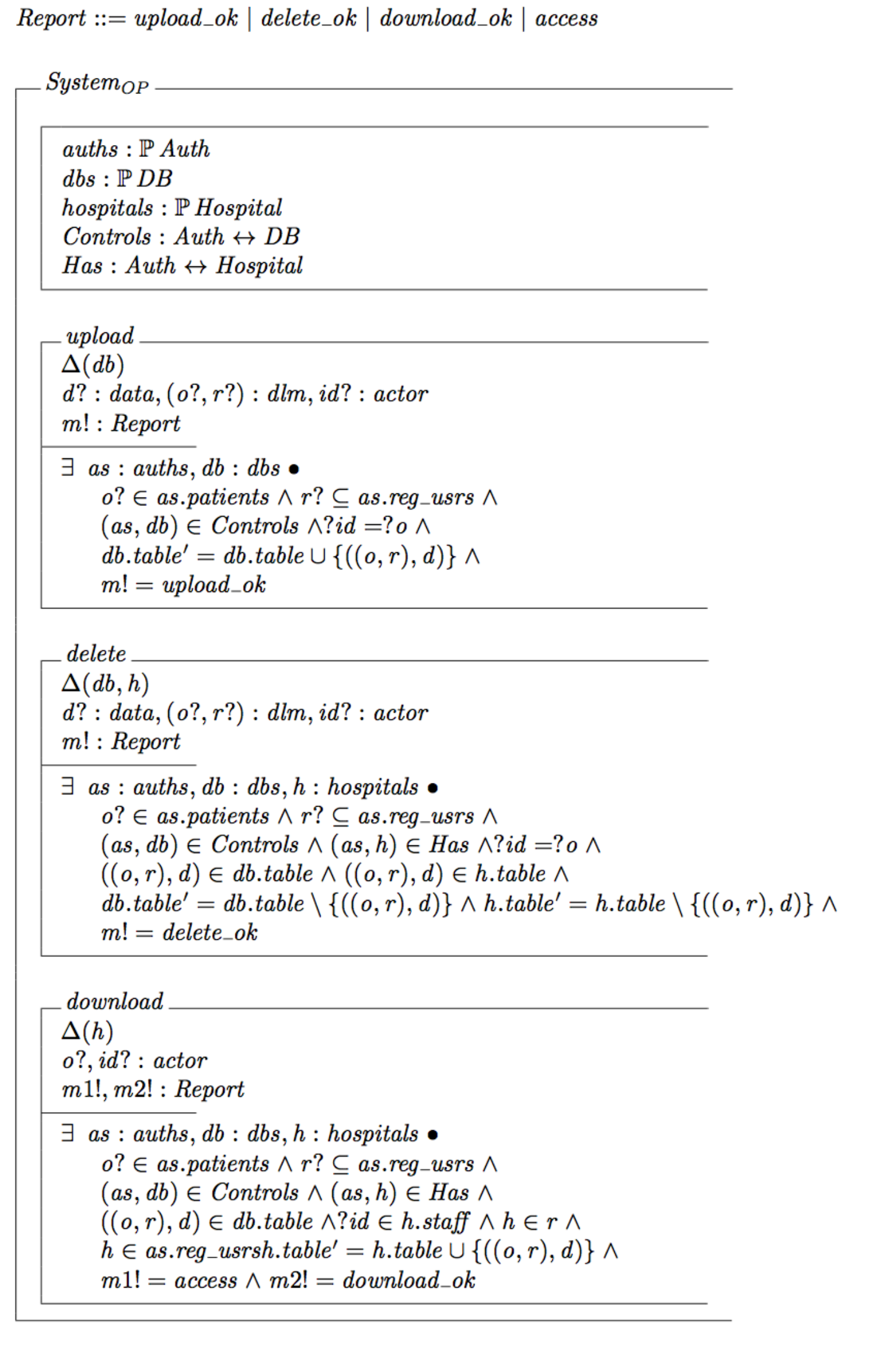}
\caption{Translation of the system model for the IoT healthcare system 
of the analysis phase}\label{fig:opoz}
\end{center}
\end{figure*}

\begin{figure*}
\begin{center}
\includegraphics[scale=.7]{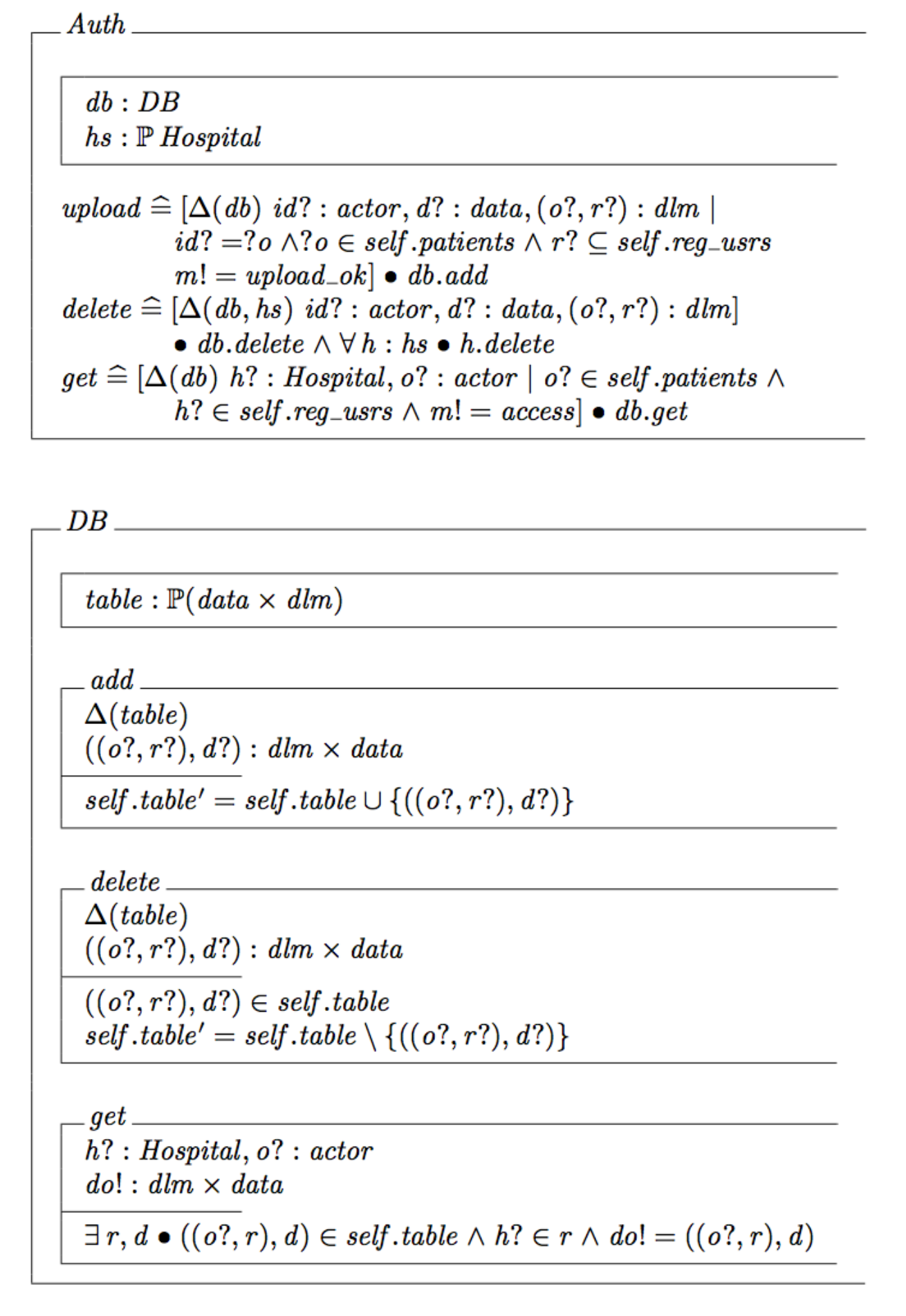}
\caption{Translation of the design model for the IoT healthcare system 
parts Auth and DB}\label{fig:objint1}
\end{center}
\end{figure*}

\begin{figure*}
\begin{center}
\includegraphics[scale=.4]{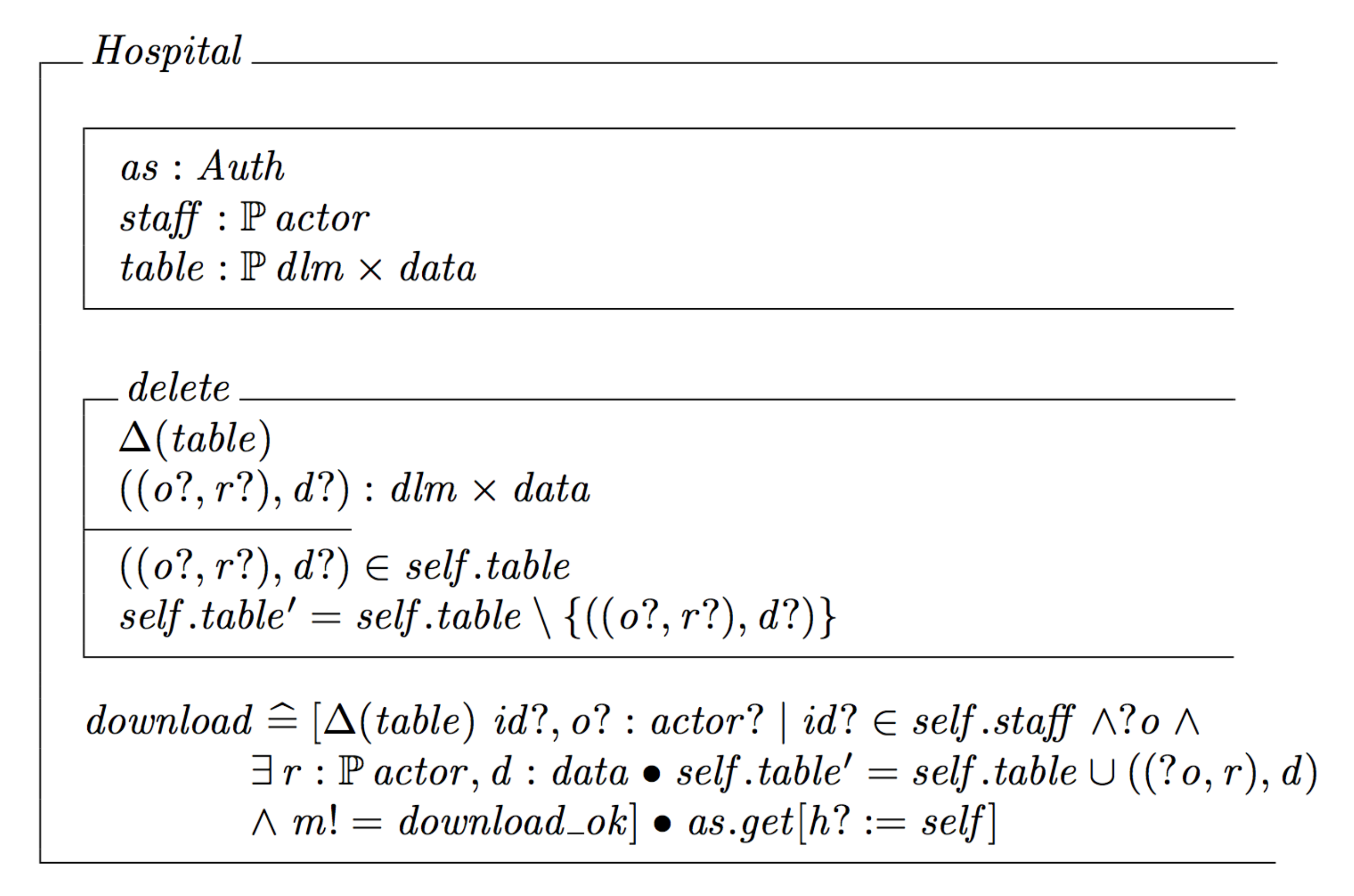}
\caption{Translation of the design model for the IoT healthcare system 
for class Hospital}\label{fig:objint2}
\end{center}
\end{figure*}

This appendix contains the output of the translation process to Object-Z.
The translation follows a schematic procedure \cite{bk:03} and thus is
created practically automatically. The translation of the system model 
of the analysis is shown in Figure \ref{fig:opoz}, the design model
derived mainly derived from the object interactions is shown for
the classes Auth and DB in Figure \ref{fig:objint1} and for Hospital
in Figure \ref{fig:objint2}.

\end{document}